# Measurement of junction conductance and proximity effect at superconductor/semiconductor junctions

Michael Vissers, Victor K. Chua, Stephanie A. Law, Smitha Vishveshwara, and James N. Eckstein

Department of Physics and
Materials Research Laboratory
University of Illinois
Urbana, IL 61801

The superconducting proximity effect has played an important role in recent work searching for Majorana modes in thin semiconductor devices. Using transport measurements to quantify the changes in the semiconductor caused by the proximity effect provides a measure of dynamical processes such as screening and scattering. However, in a two terminal measurement the resistance due to the interface conductance is in series with resistance of transport in the semiconductor. Both of these change, and it is impossible to separate them without more information. We have devised a new three terminal device that provides two resistance measurements that are sufficient to extract both the junction conductance and the two dimensional sheet resistance under the superconducting contact. We have compared junctions between Nb and InAs and Nb and 30% InGaAs all grown before being removed from the ultra high vacuum molecular beam epitaxy growth system. The most transparent junctions are to InAs, where the transmission coefficient per Landauer mode is greater than 0.6. Contacts made with ex-situ deposition are substantially more opaque. We find that for the most transparent junctions, the largest fractional change as the temperature is lowered is to the resistance of the semiconductor.

## Introduction

Recently there has been considerable interest in using the superconducting proximity effect to search for Majorana Fermion modes in semiconductor systems[1]. The possibility that such modes might exist near the ends of a superconducting contact to a high Z semiconductor nanowire was first proposed by SauLutchyn and Das Sarma[2] and subsequently refined by them and others.[3] More recently Kouwenhoven and coworkers[4] presented experimental data consistent with the emergence of zero energy modes when all of the necessary ingredients proposed in reference 1 were present, superconductivity, a magnetic field, large spin orbit interaction, a nanowire geometry and sufficient carrier doping. Of these requirements, the most difficult to guarantee and measure is the strength of the superconductive proximity coupling[5] into the semiconductor wire. It is essential to make this coupling sufficiently strong and to adjust the induced gap to match the Zeeman splitting from the applied magnetic field. This depends critically on the contact conductance. Here we present a new measuring device that has three terminals and allows direct measurement of the transmission coefficient of electron trajectories at a metallic superconductor to semiconductor interface as well as changes in the conductivity of the semiconductor layer in contact to the superconductor. We present data from two different niobium/semiconductor junctions, both made in-situ without exposing the semiconductor surface to any residual gas contamination.Under the most promising conditions the transmission factor for electrons across the junction reaches greater than 0.7. Both the choice of semiconductor and interface cleanliness are important to obtain such contacts

The resistance of a contact between a normal metal and a superconductor changes as a function of temperature for several reasons. First, low energy carriers in the normal metal have fewer final states to go to when the superconductor develops a gap. For sufficiently transmissive junctions, Andreev reflection mitigates this resistance increase. The resistance of normal layer in contact to superconductor also changes because of the induced pair correlation caused by the proximity effect, and this may also be expected to change the junction transmission as well.

The original studies of Andreev reflection[6] sidestepped this complication by limiting the contact area to smaller than the Sharvin limit. This limited the proximity effect since the contact was small, and kept the resistance of the contact large and relatively easy to measure. Proximity effect contacts, such as those in experiments searching for Majorana modes are larger than the Sharvin area, and this complicates measurement of the contact conductance.This is because a two terminal (2T) transport measurement of ajunction between a superconductor and a normal metal includes the contact resistance of the SC-N junction in series with the normal layer resistance, making it impossible to untangle each contribution to the resistance separately. All previous studies have been subject to this limitation.[7] Since the normal layer resistance is relatively large in a thin semiconductor, this complication is serious. It would be useful to know in detail how each of these quantieschange when proximity induced pairing occurs in the normal metal. To understand what the induced pair correlations do in a conductor one would like to how all the changes to the device resistance are separately manifest.

## Proximity Effect Transport Device

To separate out these two contributions to the total resistance, another measurement must be made providing independent information about the potential profile existing in the semiconducting layer below the superconductor contacts. We have devised a three terminal device shown in figure 1 that provides this extra information. A normal layer in the form of a two dimensional bar, has three superconducting contacts on top. Current is driven between the middle "injector" electrode and one of the two outside electrodes. The voltage differences between the injector and both of the outer electrodes are the two voltage measurements. In general, working back from these two measurements to extract the desired parameters characterizing the junction and the proximity effect can be handled numerically. In the case of contact to a sufficiently thin normal layer, a closed form solution can be obtained. Because of the extreme aspect ratio, transport from the superconductor into the normal layer occurs perpendicular to the junction, while most of the current flow in the normal metal flows parallel to the junction. The problem, then, effectively separates into two distributed one dimensional problems in series, transport from the superconductor into the normal layer governed by the junction conductance, $G_C$, and transport in the normal layer governed by the two dimensional resistivity (sheet resistance) $R_S$. These two quantities are experimentally found to be substantially independent of temperature above $T_C$, but to smoothly change below $T_C$. These changes are manifestations of the proximity effect.

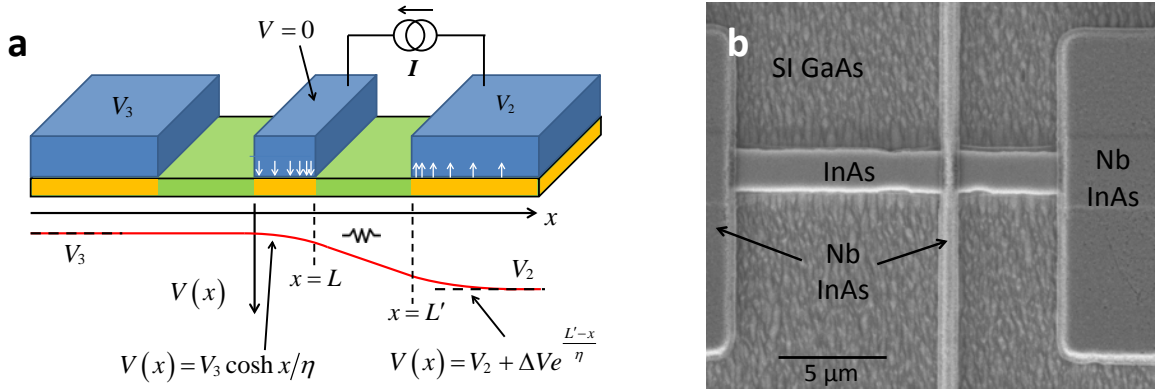

Figure 1.Panel A: A thin semiconductor bar doped with electrons colored green and orange has three highly conductive, e.g., superconducting, contacts (blue) patterned on its surface. The middle contact injects electrons from the superconductor into the thin semiconductor. The right contact extracts electrons and measures the voltage relative to the injector. The left contact measures the voltage of the semiconductor at the upstream edge of the injector. The current transfers into and out of the semiconductor over a transfer length given by $\eta = (G_C R_S)^{-1/2}$, where $G_C$ is the specific contact conductance (S/m$^2$) between the superconductor and the semiconductor, and $R_S$ is the 2D sheet resistance of the semiconductor under the superconductor. Panel B: SEM picture of finished device. The thin NbN/Nb wire is the injector, here 0.85 µm long and 2 µm wide. The Nb layer on top of the InAs film was deposited in-situ, resulting in a contact with $G_C > 10^{12}\,\mathrm{S/m^2}$ corresponding to a Landauer mode transmission of 60%. Current is extracted from the right electrode. The granular appearance of the substrate is due to etching the conducting InAs channel. The InAs lattice relaxed by generating dislocations which the etch decorated.

A superconducting finger extends from $x=0$ to $x=L$ and injects current into the normal layer below. The current is extracted from a large area contact downstream.In such a device, current is transferred from the superconductor to the semiconductor over an exponential transfer length, $\eta$, determined by the junction conductance and the semiconductor sheet resistance, as derived below. If the injector length, $L$, is long, current injection crowds near $L$ and the injected current density at $x=0$ is essentially zero. On the other hand, if $L$ is small enough current flows across the interface even at $x=0$, leading to a voltage detected by the leftmost electrode which we call $V_3$. The downstream contact extends far enough to the right in the figure that $dV/dx \to 0$, and the voltage measured by the extractor electrode is the asymptotic value in the semiconductor bar, $V_2$.We work in the linear regime and treat the transport classically. We obtain closed form expressions for the junction conductance, $G_C$, and the two dimensional sheet resistance $R_S$ of the semiconductor layer below in terms of $V_2$, $V_3$ and the resistance of the semiconductor between the injector and extractor.

When the semiconductor layer is much thinner than L, the voltage profile in the semiconductor mainly varies in-plane and can be approximated to be only a function of x, $V(x)$. The voltage difference between the injector ($V_{inj} \equiv 0$) and the semiconductor below determines the current density injected from the finger everywhere along the length of the injector, $J(x) = G_C V(x)$. This in turn determines the voltage profile in the semiconductor since $dV/dx = R_S K(x)$ where $K(x)$ is the sheet current density at x, and $R_S$ is the sheet resistance of the thin normal layer. Since $dK/dx = J(x)$, we obtain $d^2V/dx^2 = G_C R_S V$. The solution to the voltage profile in the semiconductor under the superconductor is:

$$V(x) = Ae^{x/\eta} + Be^{-x/\eta} . \qquad (1)$$

Here $\eta = \sqrt{1/G_C R_S}$ is the exponential transfer length, and the coefficients A and B must be chosen to satisfy the boundary condition that $dV/dx = 0$ at the upstream end of the injector and the downstream end of the extractor. In the case of the semiconductor under the injector this gives $V(x) = V_3 \cosh(x/\eta)$, while in the case of the extractor which is taken to be many transfer lengths long the voltage exponentially drops to $V_2$. Integrating $J(x)$ from 0 to L we obtain the total current, $I = w\eta V_3 G_C \sinh(L/\eta)$ where w is the width of the transport bar. Using this we can obtain the voltage drop from L', the start of the extraction electrode, to the asymptotic value $V_2$, since the current extracted must equal the current injected. We obtain $V(L') - V_2 = V_3 \sinh(L/\eta)$. Defining $\tilde{R}_2 = V_2 / I - R_{GAP}$ and $R_3 = V_3 / I$, where $R_{GAP}$ is the resistance of the N-layer gap between the injector and the extractor, we obtain

$$R_3 = \frac{1}{w}\sqrt{\frac{R_S}{G_C}} \frac{1}{\sinh(L/\eta)} \text{ and } \tilde{R}_2 = R_3 e^{L/\eta} .$$

Finally, we can solve for explicit expressions for $R_S$ and $G_C$:

$$R_S = \frac{w}{L} \frac{\tilde{R}_2^2 - R_3^2}{2\tilde{R}_2} \ln\frac{\tilde{R}_2}{R_3} \qquad (2)$$

$$G_C = \frac{1}{wL} \frac{2\tilde{R}_2}{\tilde{R}_2^2 - R_3^2} \ln\frac{\tilde{R}_2}{R_3} \qquad (3)$$

These results assume that the only change in transport due to the proximity effect occurs at the interface where $G_C$ may change and directly under the superconducting electrodes where $R_S$ may change. If the proximity effect is strong enough that the resistance of the N-layer between the contacted areas is changed, then this simple solution fails. This amounts to a proximity inducedchange of $R_{GAP}$, and it can be accounted for by measuring two devices with different injector lengths, $L_1$ and $L_2$. While a closed form solution is no longer possible, the information obtained by this additional measurement allows $R_{GAP}$, $G_C$ and $R_S$ to be obtained numerically.

**Experiment**

We have used Eqs 2 and 3 to investigate the temperature dependence of the junction conductance and semiconductor sheet resistance of three samples with different layer architecture to study the effect of junction transparency on the proximity effect in the semiconductor. Schematic diagrams of the layer architectures are shown in figure 2. The semiconductor heterostructures were all grown on semi-insulating GaAs substrates by molecular beam epitaxy (MBE) in a chamber that has produced two dimensional electron gas samples with mobility higher than 50 m²/Vsec at 4 K. After the semiconductor layers were grown, all samples were transferred under ultra-high vacuum conditions to a second MBE system where they were capped with 50 nm of *in-situ* deposited niobium. This is crucial for studying the intrinsic proximity effect since it results in nochemical impurity incorporation at the niobium-semiconductor junction. After being removed from the vacuum system, the wafers were introduced into a sputtering system where they were initially cleaned by ion milling to prepare the niobium surface. Then they were coated with an additional 200 nm thick film of niobium nitride which has a $T_C$ of ~13 K. The junction between Nb and NbN should be negligible compared to the Nb to semiconductor junction, since both materials are high carrier density metals. Using a combination of optical and electron beam lithography, reactive ion etching to remove Nb and wet chemical etching to remove semiconductor layers, devices like the one showed in figure 1 were made.

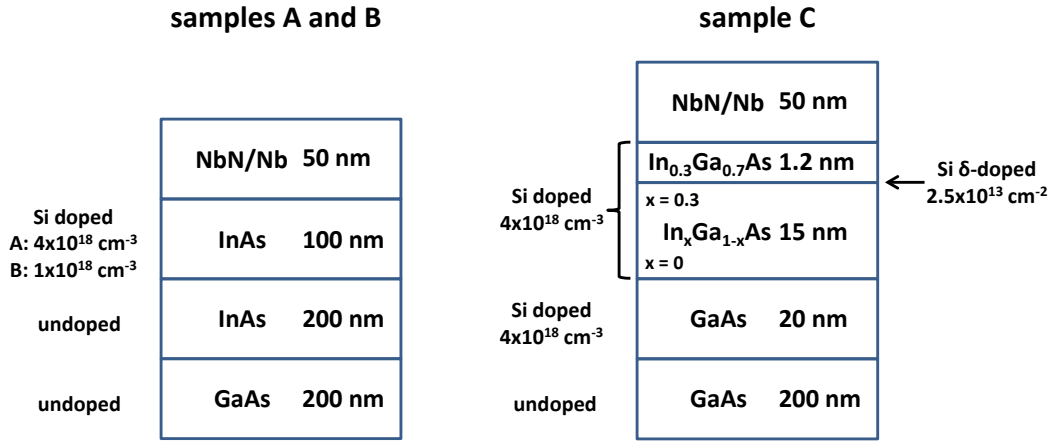

Figure 2. Epitaxial structure of the semiconductor heterostructure and the superconducting contact on top. Samples A and B differ in the doping used in the top InAs layer. The semiconductor in contact to the superconductor on top in sample C is $In_{0.3}Ga_{0.7}As$, which has a Schottky barrier of about 0.4 eV.

In samples *A* and *B*, the carrier density of the transport layer in contact to the Nb film was controlled by doping the InAs layer with Si. For comparison, two different values were chosen, $4\times10^{24}$ m$^{-3}$ for sample *A* and $1\times10^{24}$ m$^{-3}$ for sample *B*. InAs is known to have a surface accumulation layer due to Fermi level pinning above the bulk conduction band edge,[8] and this gives rise to a relatively transparent metallic contact between the InAs and Nb layers.[9] The films were grown on top of a lattice mismatched substrate. The lattice constant of InAs is 0.61 nm, while the lattice constant of GaAs is 0.56 nm. Theundoped 200 nm InAs layer accommodated dislocations that were generated due to the large lattice mismatch between GaAs and InAs. At the InAs-GaAs interface the electron diffraction pattern showed the emergence of transmission spots caused by growth of three dimensional grains. These went away during the growth of that layer. During the growth of the doped InAs layer on top, the reflection high energy electron diffraction (RHEED) pattern showed half order and third order reconstruction similar to what is observed in the growth of InAs under similar conditions on a lattice matched substrate (GaSb).

Sample C has a more complex structure. In this sample the niobium contacts a semiconductor that has a Schottky barrier, resulting in a less transparent junction. To improve the contact conductance, it is capped with a layer of $In_{0.3}Ga_{0.7}As$ which has a lower Schottky barrier than GaAs, approximately 0.4 eV.[10] The In concentration was graded from zero to 30% over a thickness of 15nm. Additionally, a heavy delta-doping layer was included 1.2 nm below the interface in an attempt to reduce the depletion depth of the Schottky barrier. These layers were grown thin enough to remain pseudomorphic and strained to the substrate.[11] Below the cap layers, 20 nm of doped GaAs also provided transport. While the doped layers were thinner than in samples A and B, the doping was higher and the sheet resistance was comparable.

The raw transport data consisting of $\tilde{R}_2(T)$ and $R_3(T)$ is shown in figures 3a and 3b. Above the $T_C$ of NbN, the wiring lines shown in figure 1b are resistive, and the analysis given above is not valid since it is spoiled by this extrinsic resistance, for example along the width of the injector. Below 13 K the wiring lines are superconducting, and eqs 2 and 3 may be used. Between 13 K and 8.5 K all three samples show weak temperature dependence, presumably due to the proximity effect between the NbN cap layer and the Nb contact layer. Below ~8.5 K the Nb is superconducting, and the proximity effect between the Nb and the semiconductor leads to significant changes in the measured resistances. Transport in similar semiconductor layers without Nb coating was also measured to quantify the nature of the material in which the proximity effect is studied. In this case the resistance versus temperature rose less than 2% below 20 K indicating degenerate metallic behavior. Therefore, the changes seen in the data below 13 K can be attributed to the effect of superconductivity coupled to the semiconductors.

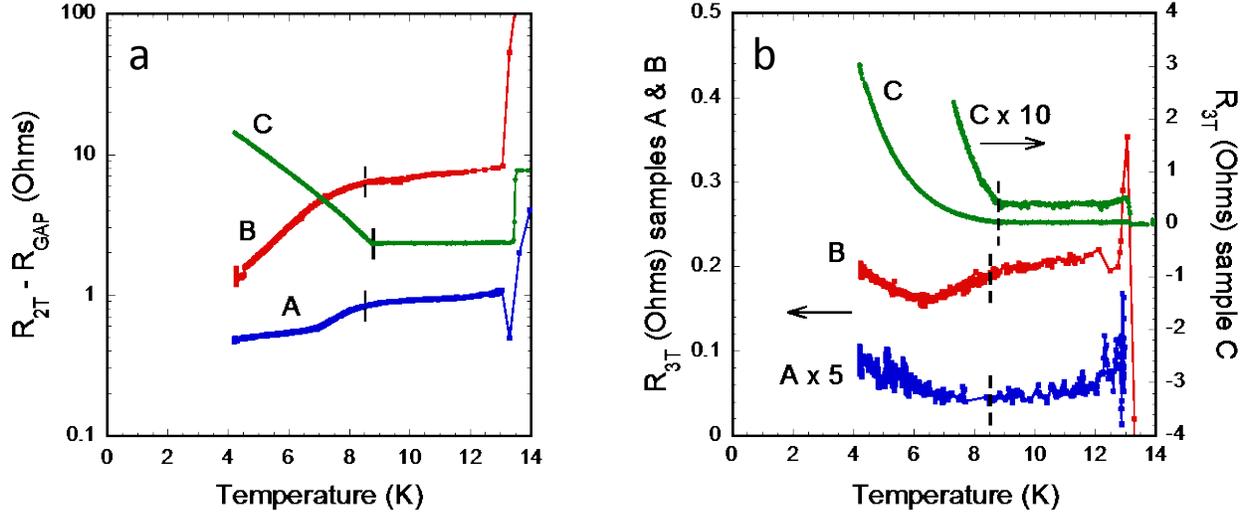

Figure 3. Resistance in two and three terminal measurements for all three samples. Panel a shows the two terminal resistance with the gap resistance subtracted $\tilde{R}_2 = V_2/I - R_{GAP}$, and panel b shows the three terminal resistance. The behavior of samples A and B is similar, but sample C is distinctly different. The dashed lines indicate the temperature at which the 50 nm niobium film becomes superconducting. The NbN superconducting transition occurs at ~13 K.

Comparing samples *A* and *B*, the higher carrier density in sample *A* leads to a more conductive sample, and this shows up in the relative resistance values measured. The data show qualitatively similar behavior for the two samples. Panel 3a shows the two terminal resistance, with the resistance of the semiconductor in the gap between the two terminal contacts subtracted out. The subtracted two terminal resistance, $\tilde{R}_2 = R_2 - R_{GAP}$, drops substantially when the niobium film becomes superconducting, by a factor of 1.8 for sample A and 4.7 for sample B. There are two reasons this may occur. First, if the contact is transparent, the junction transport becomes dominated by Andreev reflection, which at a perfectly transparent interface leads to an increase in the junction conductance by a factor of two.[12] Second, if the proximity effect is substantial, the pair correlations may lead to reduced semiconductor resistivity. On the other hand, the three terminal resistances in both sample A and B rise as T→0. This depends on the both $G_C$ and $R_S$ in a non-trivial manner and provides the second piece of information to obtain them as shown in eqs 2 and 3.

Sample *C* exhibits different behavior. Both $\tilde{R}_2$ and $R_3$ rise suddenly at 8.5 K when the Nb film becomes superconducting. Because the contact is expected to be less transparent due to the Schottky barrier at the interface, it is likely that the junction behaves like a tunnel contact. Since the resistance of the semiconductor without the superconducting contact is temperature independent, the increased resistance in both of these measurements must be largely due to tunneling through the Schottky barrier into the temperature dependent quasiparticle density of states in the Nb layer.

Using eqs 2 and 3 the junction conductance and sheet resistance as a function of temperature can be extracted from the two resistance measurements for each device, and these are shown in figures 4a and b. Samples *A* and *B* with contact between Nb and InAs show similar behavior, while sample *C* which has a Schottky barrier contact behaves essentially oppositely. The same distinction between the two types of samples is true for the semiconductor sheet resistance as a function of temperature as well. $R_S$ for both samples *A* and *B* decrease substantially when the niobium layer becomes superconducting, while sample C it actually increases.

The value of the junction conductance and the shapes of the $G_C(T)$ curves provide quantitative information about the proximity effect that can be compared with theory. In samples A and B the ratio of $G_C(0K)/G_C(T_C)$ gives a measure of this. For sample *A* it is 1.65, while for sample B it is 2.89. Both of these numbers are greater than one, and reflect the increase in conductance caused by Andreev reflection. In this case the conditions of the sample and measurement are different from those found in point contact spectroscopy, as described by Blonder, Tinkham and Klapwijk. Here the contact is large area, and the pair correlations induced in the normal layer should be added to the boundary match equations at the interface. The large increase in junction conductance, especially in sample B at least indicates a similar physical process is at work. Additionally, the value of junction conductance just below the TC of the NbN layer can be used to derive the transmission coefficient for each quantum channel. The contact conductance can be written as

$$G_C = \frac{2e^2}{h}\left(\frac{2}{\lambda_F}\right)^2 |T|^2. \tag{4}$$

Here $\lambda_F$ is the Fermi wavelength and T is the amplitude transmission coefficient across the junction. The junction conductance at 13 K for sample A is $1.3 \times 10^{12}$ S/m$^2$ and for sample B is $4.5 \times 10^{11}$ S/m$^2$. Recalling that for simple parabolic Fermi systems with an energy independent effective mass $k_F = (3\pi^2 n)^{1/3}$ we obtain $|T_A|^2 = 0.69$ and $|T_B|^2 = 0.60$. Even though the actual junction conductance at $T_C$ differs by a factor of almost three, they have nearly the same transparency. This indicates that the chemical reaction forming the junction between the Nb and the InAs is reproducible when carried out in this pristine manner. The numbers obtained are remarkably high considering the difference in the two materials. The discussion of junction barrier by BTK concludes that even when there is no chemical barrier there will still be an effective barrier contribution due to differences in Fermi velocity at the interface.[6] It turns out in this case that the larger carrier density in Nb is largely offset by the small effective mass in InAs. For Nb the average $v_F$ is $1.4 \times 10^6$ m/s, while for InAs it ranges between 1.2 to $1.9 \times 10^6$ m/s. According to the result in ref. X, an effective barrier, $Z_{eff}$, exists at the boundary for an otherwise perfect junction due to mismatched Fermi velocity. Using these numbers we find that $Z_{eff} \leq 0.15$, which even for simple Andreev reflection results in a reduction of the zero temperature conductance by approximately 10%.

The temperature dependence of the junction conductance measured on sample *C*, which has a Schottky barrier at the interface, is accurately fit by the functional form of the temperature dependence of the Nb quasiparticle density as shown in the inset to figure 4a. Here the relative BCS quasiparticle density with a gap of 1.18 mV overlays the measured junction conductance, indicating that the temperature dependence in the measured junction conductance is entirely due to freezing out quasiparticles. The conductance at $T_C$ is $2 \times 10^{11}$ S/m$^2$. An estimate for the Fermi wavelength at the interface can be obtained combining the bulk doping with the delta doping 1.2 nm below and estimating the electron density is spread over a thickness of order three times the Thomas-Fermi screening length which is about 3.5 nm. This can be used to obtain $\lambda_F$ and with this we compute $|T_C|^2 = 0.032$, squarely in the tunneling regime.

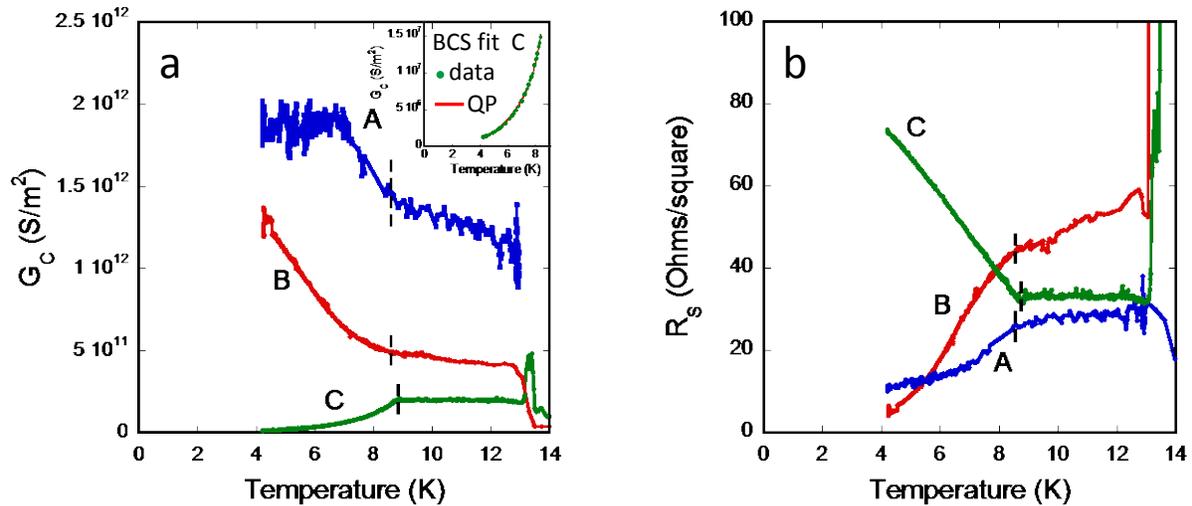

Figure 4. Panel A shows the junction conductance between the Nb superconductor and the underlying semiconductor. The inset shows that the reduction in junction conductance seen in sample C is explained by tunneling into a BCS density of states. Panel B shows the semiconductor sheet resistance. The substantial reduction in RS seen in samples A and B which have transparent interfaces is caused by proximity induced pair correlations in the semiconductor. The rise in sheet resistance seen in sample C may be caused by scattering from the delta doping layer just below the interface. It appears that superconducting correlations contribute to the scattering which is unexplained.

The sheet resistance of the semiconductor below the Nb is also changed by the proximity effect, and this is shown in figure 4b. While for $G_C(T)$ Andreev reflection provides an enhancement for transparent junctions independent of whether pair correlations exist in the semiconductor, for $R_S(T)$ any change observed from the value at $T_C$ is directly traceable to the existence and magnitude of pair correlations. (Without the Nb contact, the RS of these materials is temperature independent over this temperature range.) Thus, this result can be used to unambiguously quantify the proximity effect. The sheet resistance for samples $A$ and $B$ which are transparent drops sharply and substantially after the Nb layer becomes superconducting. The reduction in $R_S$ seen between 13 K and 8.5 K appears to be due to pairs diffusing from the NbN layer. We conclude that a very important consequence of the proximity effect is to enhance normal transport in the semiconductor just adjacent to the junction. Apparently pair correlations lead to extra conductance, similar to paraconductivity seen above TC in bulk superconductors. Here the pair correlations do not arise from an internal pairing potential, but rather from the neighboring superconductor. It is also interesting to note that as in the $G_C$ data, the extent of the change induced by the proximity effect is larger for sample $B$ than it is for sample $A$. This may be caused by crystal quality, since the higher doping in sample $A$ also must lead to a shorter mean free path.

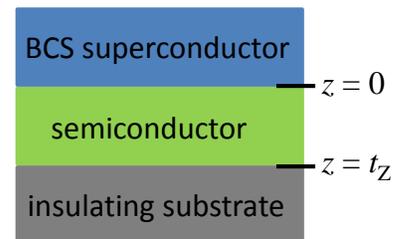

Figure 5. A one dimensional superconductor – normal – insulator (SN) junction. Here $t_Z$ denotes the thickness of the N-layer which for devices A,B,C are 100nm,100nm and 35nm respectively.

**Quasiclassical Theory**

The quasiclassical theoretical framework has proven to be very successful in capturing superconductivity-normal metal proximity physics in micro-devices[13,11] and vastly generalizes the transmission matrix or Landauer transport theory implicit in the original BTK analysis and subsequent refinements.[14] We refer the reader to the excellent review by Klapwijk[15] for the relationships between these different approaches and to references 9 and 16 for reviews of the quasiclassical method. The power and utility of the quasiclassical formalism lies in its ability to not only map out the spatially varying electronic properties in inhomogeneous systems with normal and superconducting "reservoirs", but also to incorporate impurity scattering, non-equilibrium situations and macroscopic superconducting phase variations. As a simplification we consider a one-dimensional geometry and ask the question of how far the proximity effect extends below the SN interface and affects the electronic states in the semiconductor.

The doping levels in the semiconductor N-layers motivates the use of the Usadel equations[9,12] for quasiclassical Green's functions which operate in the diffusive transport regime. The Usadel equations which describe superconductivity as a function of position and energy contain all the information necessary to calculate the local spectroscopic and transport properties for both the superconductor as well as the proximity coupled normal metal. It is represented by a complex angle $\theta_{N,S}(\vec{r},E)$ parameterizing the Fermi-velocity averaged (s-wave component) of the retarded quasiclassical Green's function in the N and S layers respectively.

The geometry of our model is shown in figure 5 and we take the S and N layers of figure 5 to be both diffusive metals when in their respective normal metallic phases. They are characterized by bulk normal state conductivities $\sigma_{S,N}$ and quasiparticle diffusion constants $D_{N,S}$. These are parameters of our model which we determine from transport measurements of our devices above $Tc$. The description of the proximity effect is essentially one dimensional in our problem so that $\theta_{N,S}$ is a function only of $z$ the distance from the junction and $E$ the energy measured relative to the Fermi-energy. $\theta_{N,S}$ solves a nonlinear second order differential equation which is Usadel's equation for the retarded quasiclassical Green's function

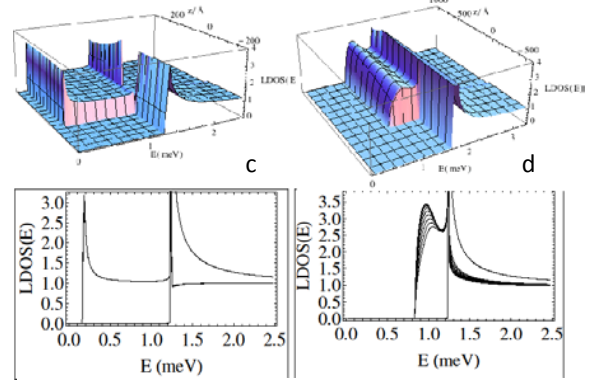

Figure 6. The normalized local density of states (LDOS) computed from the solutions of Eqns. (5,6) for parameters derived from device A and C or different characteristics and thicknesses. The order parameter in the S-layer is $\Delta(T) =$ 1.23meV at T=4K, the junction interface is exactly at z=0 where N, z>0 and S, z<0. **(a)** Device C for the not very transparent device of with extracted Gc = $2 \times 10^{11}$S/m$^2$. **(b)** Device A for the not very transparent device of with extracted Gc = $2 \times 10^{12}$S/m$^2$. **(c)** Traces of differing depths taken from **(a)** showing the small "minigap" of about 0.1meV. **(d)** Traces of differing depths taken from **(b)** displaying a much larger "minigap" of about 0.8meV.

$$\hbar D_N \partial_z^2 \theta_N(z,E) + 2iE \sinh \theta_N(z,E) = 0, \quad z > 0 \tag{5}$$

$$\hbar D_S \partial_z^2 \theta_S(z,E) + 2iE \sinh \theta_S(z,E) - 2i\Delta(T) \cosh \theta_S(z,E) = 0, \quad z < 0 \tag{6}$$

where $\Delta(T)$ is the temperature dependent superconducting order parameter in S. These equation are not decoupled but are to be solved with the following Kupriyanov- Lukichev[17] boundary conditions at the interface for opaque barriers for all energies $E$,

$$\sigma_N \frac{\partial \theta_N}{\partial z}\bigg|_{z=0^+} = \sigma_S \frac{\partial \theta_S}{\partial z}\bigg|_{z=0^-} = G_C \sinh\big(\theta_N(z=0^+) - \theta_S(z=0^-)\big) \tag{7}$$

where $G_C$ is specific contact conductance. The parameter $G_C$ is in fact the same contact conductance that is meant to be extracted from the three terminal expression (3) prior by the proximity effect taking its hold on the N-layer.

Measurable physical quantities predicted by the theory can then be computed from $\theta_{N,S}(E,z)$. One such quantity that will concern us is the *normalized* local density quasiparticle density of states $\text{LDOS}(z,E) = \text{Re}[\cosh\theta(z,E)]$. Multiplied by the normal state Fermi-energy density of states yields the quasiparticle spectral density of states modified by the proximity effect. In the limit of a BCS superconductor, the LDOS is *zero* in the energies $|E| < \Delta(T)$. The proximity effect is manifest in the N-layer by the appearance of an LDOS gap as well. However the size of this spectral gap is diminished compared to $\Delta(T)$, and as we show from theoretical computations in figure 6, very sensitive to the contactance conductance $G_C$. This corroborates the data from transport measurements and highlights the sensitivity of the proximity effect on the junction conductance $G_C$.

**Acknowledgements**

"This material is based upon work supported by the U.S. Department of Energy, Division of Materials Sciences under Award No. DE-FG02-07ER46453, through the Frederick Seitz Materials Research Laboratory at the University of Illinois at Urbana-Champaign."